\documentclass[aps,pre,twocolumn,floatfix,showpacs,amsmath,10pt]{revtex4-1}
\usepackage{amssymb,graphics}
\usepackage{setspace} 
\usepackage{graphicx}
\usepackage{hyperref}
\usepackage{xcolor}

\newcommand{\showfigures}[1]{{#1}} 

\newcommand{\DP}[2]{\ensuremath{\frac{\partial{#1}}{\partial{#2}}}}

\newcommand{\DT}[2]{\ensuremath{\frac{\mathrm{d}{#1}}{\mathrm{d}{#2}}}}

\newcommand{\kf}{K_{\mathrm{F}}}
\newcommand{\kr}{K_{\rho}}
\newcommand{\rhoe}{\rho_{\mathrm{e}}}
\newcommand{\frho}{\psi_{\rho}}
\newcommand{\fp}{\varphi_{p}}

\newcommand{\del}{\partial}

\begin{document}
\title{Emergence of epithelial cell density waves}
\author{Shunsuke Yabunaka$^1$} 
\email{yabunaka@scphys.kyoto-u.ac.jp}
\author{Philippe Marcq$^2$}
\email{philippe.marcq@curie.fr}

\affiliation{$^1$ Fukui Institute for Fundamental Chemistry, Kyoto University, Kyoto, Japan}
\affiliation{$^2$ Laboratoire Physico Chimie Curie,
Institut Curie, Universit\'e Marie et Pierre Curie, CNRS UMR 168, 
26 rue d'Ulm, F-75248 Paris Cedex 05, France}

\date{August 22, 2017}

\begin{abstract}
Epithelial cell monolayers exhibit traveling mechanical waves. 
We rationalize this observation thanks to a hydrodynamic
description of the monolayer as a compressible, active and polar material.
We show that propagating waves of the cell density, polarity, velocity and 
stress fields may be due to a Hopf bifurcation occurring above threshold 
values of active coupling coefficients.

\end{abstract}

\maketitle

\section{Introduction}
\label{sec:intro}

Pattern-forming instabilities \cite{Cross1993} play a pivotal role 
in the morphogenesis and spatio-temporal organization of living organisms. 
Paradigmatic examples include the morphogen patterns of developing 
embryos \cite{Meinhardt2008} and the cooperative behaviour of large 
assemblies of microorganisms \cite{BenJacob2000}, for instance in 
bacterial colonies.
Nonlinear oscillators contribute to our understanding of phenomena as 
diverse as genetic clocks \cite{Kruse2011} or calcium signaling 
\cite{Falcke2003}. Most processes occurring at the scale of a single 
cell are strongly dissipative and driven by the consumption of chemical
fuel. A recognition of this basic fact has led in the past few decades 
to a strong cross-fertilization of concepts and methods between 
non-equilibrium physics and cell biology \cite{Karsenti2008,Beta2017}.

At the scale of the tissue, Serra-Picamal \textit{et al.}
\cite{Serra-Picamal2012} discovered that the expansion into free space 
of  Madin-Darby canine kidney (MDCK) epithelial cell monolayers is 
accompanied by traveling mechanical waves of the velocity, strain rate, 
traction force and stress fields. This observation has since been 
confirmed independently by others, in the same geometry 
\cite{Tlili2015,Tlili2016}, as well as in confined domains 
\cite{Notbohm2016}. To the best of our knowledge, the biological 
function of the epithelial waves remains unknown. 
\emph{In vitro} epithelial cell monolayers are a robust model system 
for epithelial tissues, amenable to controlled experiments
in various geometries and conditions
\cite{Poujade2007,Vedula2012,Cochet-Escartin2014,Bazellieres2015}.
Monolayer expansion has been studied theoretically by several authors
\cite{Lee2011a,Koepf2013a,Recho2016,Blanch2017a}, and shown to
be conducive to spatio-temporally disordered dynamical regimes
\cite{Lee2011b,Koepf2013b}.
The first theoretical account of the mechanical waves given in
\cite{Serra-Picamal2012} necessitated a nonlinear cell rheology.
More recent models involve either a linear elastic
\cite{Banerjee2015,Notbohm2016,Tlili2016} or a  viscous \cite{Blanch2017b} 
rheology, as well as couplings of the mechanical stress to additional 
fields such as the myosin density \cite{Banerjee2015,Notbohm2016} 
or the cell polarity \cite{Blanch2017b,Tlili2016}.
Although traveling waves of the cell density field have been 
observed during epithelial spreading \cite{Tlili2016}, 
the  evolution equation of the cell density has not been 
included in existing models. This is all the more surprising that
the cell density field of epithelial cell monolayers 
is known to fluctuate in both time and space, both \emph{in vitro} 
\cite{Zehnder2015,Zehnder2015a} and \emph{in vivo} \cite{Guirao2015}:
the corresponding  epithelial flows are thus \emph{compressible}.

In this work, we formulate and justify a minimal physical
model of a cohesive tissue able to sustain traveling mechanical waves.
Noting that the waves occur in the absence of cell division
\cite{Tlili2015}, we defer a rigorous treatment of cell proliferation 
to a separate work \cite{Yabunaka2016} and treat the
cell density field as a conserved quantity.
Importantly, waves are suppressed by inhibitors of contractility,
such as blebbistatin \cite{Serra-Picamal2012}. We therefore consider 
tissue activity as a necessary ingredient.
Experiments have shown that inhibitors of polarity also suppress the waves 
\cite{Tlili2016}: as a consequence we include tissue polarity in our
description. Within the framework of linear nonequilibrium thermodynamics
\cite{Chaikin2000}, we write the constitutive equations for 
a one-dimensional, compressible, polar and active material,
including lowest-order nonlinear active terms 
\cite{Kruse2005,Marchetti2013} that involve the cell density field.
We show that Hopf bifurcations occur beyond threshold values of active 
parameters, leading to traveling waves for the relevant mechanical fields
(cell density, velocity and polarity), and thus also for the tissue 
traction force and stress fields. Inhibition assays of active and polar
behaviour are mimicked by numerical simulations of our model equations. 
We compare our minimal model  with competing formulations, and discuss 
the robustness of our findings  to the inclusion of additional terms.

\section{Model}
\label{sec:model}

Since relevant experiments were performed in a quasi-one-dimensional
geometry \cite{Serra-Picamal2012,Tlili2015,Notbohm2016}, we focus on 
the one dimensional case. We enforce periodic boundary conditions in a 
system of constant spatial extension $L$, 
with space coordinate $x$ and time $t$.

\subsection{Conservation equations}
\label{sec:model:conservation}

The conservation equation for cell number reads:
\begin{equation}
\label{eq:cons:matter}
\del_t \rho + \del_x \left(\rho v\right) = 0 \,,
\end{equation}
where $\rho(x,t)$ and $v(x,t)$ respectively denote the 
cell number density and velocity fields.
Denoting $\sigma(x,t)$ the (internal) tissue stress field
and $t^{\mathrm ext}(x,t)$ the (external) force exerted by the substrate 
on the tissue, the conservation of linear momentum is expressed as:
\begin{equation}
\label{eq:fbal}
\del_x \sigma =  - t^{\mathrm ext} \,.
\end{equation}
Motile cells within the tissue are polar, with a well-defined
front and rear: coarse-graining over a suitable mesoscopic scale
the cell polarity yields the tissue polarity field $p(x,t)$.
The external force is assumed to be due both to fluid friction, with  a
friction coefficient $\xi$, and to active motility forces oriented along 
the polarity $p(x,t)$, with an amplitude $t_{\mathrm{a}}$ 
\cite{Banerjee2015,Notbohm2016,Tlili2016,Blanch2017b}:
\begin{equation}
\label{eq:friction}
t^{\mathrm ext} = - \xi \, v + t_{\mathrm{a}} \, p \,.  
\end{equation}

\subsection{Thermodynamics}
\label{sec:model:thermo}

Including the spatial gradients of the cell density and
polarity fields, we assume that the total free energy of the 
cell monolayer reads
\begin{equation*}
  \label{eq:freeenergy}
F=\int \mathrm{d}x \,\,
f(\rho, \partial_x \rho, \partial^2_x \rho, p, \partial_x p, \partial^2_x p) 
\,, 
\end{equation*}
where integration is performed over the domain of size $L$, and 
$f$ is the free energy density.
Here and below a dot denotes the total derivative, as in 
$\dot{f} = \partial_t f + v \, \partial_x f$.
We evaluate the free energy variation rate
$$
  \dot{F}  = \DT{}{t} \int \mathrm{d}x \,   f  
 = \int \mathrm{d}x \, \partial_t  f 
 = \int \mathrm{d}x \, \left( \dot{f} - v \partial_x f \right)
 = \int \mathrm{d}x \, \left( \dot{f} + f \partial_x v \right)
$$
following an integration by parts.
The chain rule yields
\begin{align*}
\dot{f} =& \DT{\rho}{t} \DP{f}{\rho}
+ \DT{}{t}\left(\partial_x \rho\right) \, \DP{f}{(\partial_x \rho)}
+ \DT{}{t}\left(\partial^2_x \rho\right) \, \DP{f}{(\partial^2_x \rho)}\\
&+ \DT{p}{t} \DP{f}{p}
+ \DT{}{t}\left(\partial_x p\right) \, \DP{f}{(\partial_x p)}
+ \DT{}{t}\left(\partial^2_x p\right) \, \DP{f}{(\partial^2_x p)} \,.
\end{align*}
Using \eqref{eq:friction}, and the periodic boundary conditions,
the power of external forces done on the monolayer reads
\begin{equation*}
  \label{eq:power}
  \Pi = \int \mathrm{d}x  \, v\left( - \xi v + t_{\mathrm{a}} p \right) 
= - \int \mathrm{d}x \,  v \, \partial_x\sigma
=  \int \mathrm{d}x \,  \sigma \, \partial_xv \,.
\end{equation*}
For isothermal transformations at temperature $T$, we obtain
$$
  T \Sigma = \Pi - \dot{F}  + \int \mathrm{d}x \,r \Delta \mu\,,
$$
where $\Sigma$ denotes the entropy production rate, and 
we included an additional (chemical) active term \cite{Kruse2005},
with $r$ and $\Delta \mu$ respectively the reaction rate
and chemical potential difference of nucleotide hydrolysis.
Following \cite{Kruse2005,Marchetti2013}, we lump under the term 
``activity'' all cell processes driven by nucleotide hydrolysis, 
which are relevant for cell and tissue mechanics. This includes 
motor-driven contractility as well as actin polymerization.

Noting that
\begin{align}
\DT{}{t}\left(\partial_x p\right) & = \partial_x \DT{p}{t} 
- (\partial_x p) (\partial_x v)  \nonumber\\
\DT{}{t}\left(\partial^2_x p\right) & = \partial^2_x \DT{p}{t} 
- (\partial_x p) (\partial^2_x v) 
- 2 (\partial^2_x p) (\partial_x v) 
 \nonumber
\end{align}
(and analogous equations concerning the density field),
integrations by part yield
\begin{equation}
  \label{eq:dissipation}
   T \Sigma = \int \mathrm{d}x \, \left(
\left( \sigma + \pi \right) \; \partial_x v 
      + \dot{p} \; h + r \; \Delta \mu \right) \,,
\end{equation}
with the following definitions of the molecular field $h$, 
conjugate to the polarity:
\begin{equation}
  \label{eq:def:h}
h = - \DP{f}{p} + \partial_x \left( \DP{f}{(\partial_xp)} \right)
- \partial^2_x \left( \DP{f}{(\partial^2_xp)} \right) \,,
  \end{equation}
and of the pressure $\pi$, conjugate to the density
\begin{align}
  \nonumber
\pi =&  - f +  \rho \left[ \DP{f}{\rho} 
- \partial_x \left(  \DP{f}{(\partial_x \rho)} \right)
+  \partial^2_x \left(  \DP{f}{(\partial^2_x \rho)} \right)
\right]
\\
\nonumber
&+ (\partial_x\rho) \, \left[  \DP{f}{(\partial_x \rho)}
- \partial_x \left( \DP{f}{(\partial^2_x \rho)}\right)  \right]
+ (\partial^2_xp) \,  \DP{f}{(\partial^2_x \rho)} 
\\
&+ (\partial_xp) \, \left[  \DP{f}{(\partial_xp)}
- \partial_x \left( \DP{f}{(\partial^2_xp)}\right)  \right]
+ (\partial^2_xp) \,  \DP{f}{(\partial^2_xp)} \,.
\label{eq:def:p}
\end{align}

\subsection{Constitutive equations}
\label{sec:model:consteq}

From \eqref{eq:dissipation}, we identify the following 
conjugate flux-force pairs 
\begin{eqnarray} 
\nonumber
 \mathrm{Flux} &\leftrightarrow& \mathrm{Force} \\
\nonumber
\sigma + \pi  &\leftrightarrow& \partial_x v \\
\nonumber
\dot{p} &\leftrightarrow& h \\
\nonumber
r &\leftrightarrow& \Delta \mu \,.
\end{eqnarray}
Following a standard procedure, the constitutive equations read
\begin{eqnarray}
  \label{eq:consteqsig}
  \sigma  +  \pi  &=&  \eta \; \partial_x v 
           + \sigma_{\mathrm{a}}  + \gamma_{\mathrm{a}} \, \rho \\
  \label{eq:consteqp}
  \dot{p} &=& \Gamma_{\mathrm{p}} \; h 
- \lambda  \; p \, \partial_x v 
- \alpha_{\mathrm{a}}  \; p \, \partial_x p
+ \delta_{\mathrm{a}}  \; \partial_x \rho
\end{eqnarray}
omitting the equation expressing as a function of thermodynamic forces
the field $r$. 
The dynamic viscosity $\eta$ and the kinetic coefficient
$\Gamma_{\mathrm{p}}$ are positive. 
We include lowest-order (nonlinear)
active coupling coefficients, denoted by the index $a$, and 
proportional to $\Delta \mu$, as in $\gamma_{\mathrm{a}} = \gamma \, \Delta \mu$.
A contractile material is characterized by a positive
active stress $\sigma_{\mathrm{a}}$, which  may ensure, for large enough 
positive values, that the cell monolayer is everywhere under tension, 
as observed experimentally \cite{Serra-Picamal2012}.
Since $r$ may be proportional to density, active coupling 
coefficients may also be proportional to $\rho$. This justifies the
active term $\gamma_{\mathrm{a}} \, \rho$ in 
\eqref{eq:consteqsig}.
The invariance properties of polar media allow for an active coupling between 
polarity and density gradient, with a coefficient $\delta_{\mathrm{a}}$
in \eqref{eq:consteqp}.
Two advection terms with coupling coefficients 
$\alpha_{\mathrm{a}}$ and $\lambda$, of arbitrary signs,
are included in Eq.~(\ref{eq:consteqp}), in agreement with
earlier studies of active liquid crystals \cite{Giomi2012}.

\subsection{Free energy density}
\label{sec:model:free:energy}

In our minimal description, we define the free energy density as
\begin{equation}
\label{eq:def:f}
  f =   \frho(\rho) +   \fp(p) 
  + \frac{\nu}{2} \, \left( \partial^2_x \rho \right)^2 
\end{equation}
(see Sec.~\ref{sec:disc:add} for possible additional terms).
The functions $\frho(\rho)$ and $\fp(p)$ depend respectively on density 
and polarity only. 
We include a stabilizing ($\nu \ge 0$) higher-order density gradient term 
$\frac{\nu}{2} \, \left( \partial^2_x \rho \right)^2$. It is akin
to the higher-order displacement gradient terms used in
second gradient elasticity \cite{Mindlin1965,Ciarletta2012}. 
We obtain the conjugate fields from 
(\ref{eq:def:h}-\ref{eq:def:p}):
\begin{eqnarray}
  \label{eq:val:h}
h &=& - \fp'(p) \\
  \label{eq:val:pi}
\pi &=& -f + \rho \left[ \frho'(\rho) +  \nu \, \partial^4_x \rho \right]
+ \nu \left[  (\partial^2_x \rho)^2  
- (\partial_x \rho) \, \left(  \partial^3_x \rho \right)
\right] \,.
\end{eqnarray}

In the following, we expand the free energy density $f$ in the vicinity
of a state characterized by a finite polarity $p_0$ and a 
finite density $\rhoe$. We thus define
\begin{equation}
\label{eq:def:fp}
\fp(p) = - \frac{a_2}{2} \; p^2 + \frac{a_4}{4} \; p^4 \,,
\end{equation}
where $a_2$, $a_4$ are positive coefficients and
\begin{equation}
\label{eq:def:frho}
\frho(\rho) =  \frac{1}{2 K} \; 
\left( \frac{\rho - \rhoe}{\rhoe} \right)^2 
 + \frac{1}{4 K_4} \; \left( \frac{\rho - \rhoe}{\rhoe} \right)^4\,,
\end{equation}
where $K \ge 0$ is the compressibility modulus,
and $\rhoe$ is a reference density. 
The quartic term, with coefficient $K_4 \ge 0$, is included for 
numerical stability. It makes the density-dependent potential 
steeper, but does not alter the linear stability analysis.

\begin{figure}[!t]
\centering
\showfigures{
(a)\includegraphics[scale=0.5]{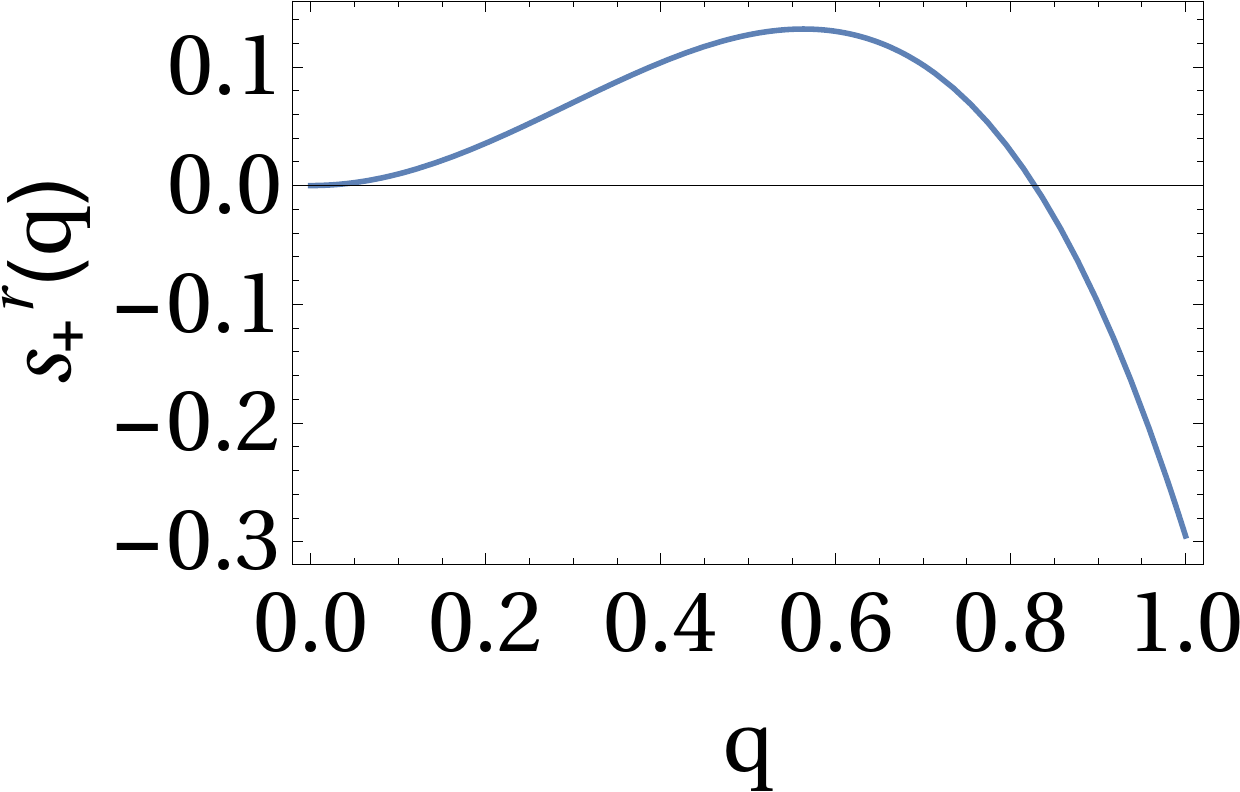}
\hfill
(b)\includegraphics[scale=0.5]{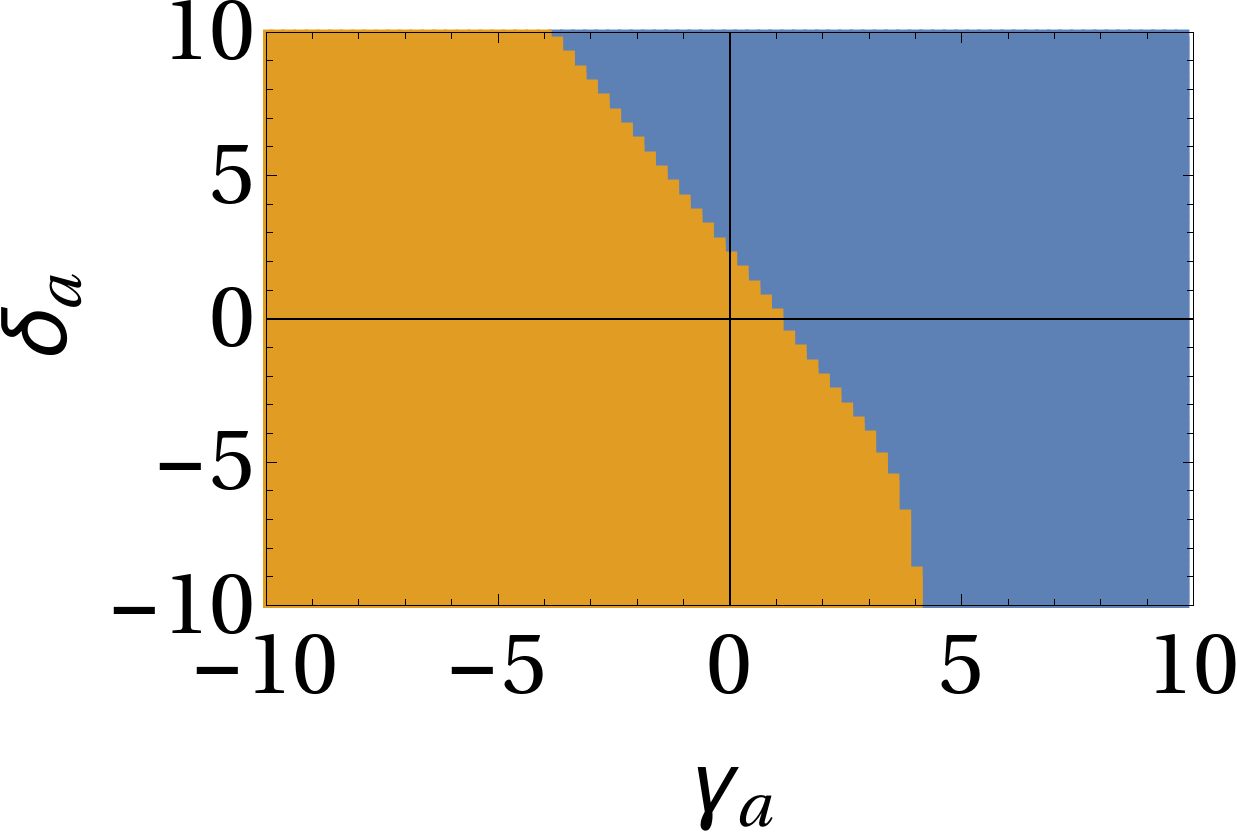}
}
\caption{\label{fig:all:one}
\textbf{Linear stability analysis.}
(a) Plot of the real part of the largest growth rate $s_{+}^r(q)$
as a function of wave number $q$.
All parameters are set equal to $1$ (including
$\gamma_{\mathrm{a}}$ and $\delta_{\mathrm{a}}$), except $p_0 = 0.5$, $\xi = 0.5$. 
We find $q_{\mathrm{max}} = \mathrm{argmax}_q(s_{+}^r(q)) = 0.56$, 
with a wave velocity $c =  s_{+}^i(q_{\mathrm{max}})/q_{\mathrm{max}} = 1.07$
greater than the mean tissue velocity $v_0 = t_{\mathrm{a}} p_0/\xi = 1$.
We checked that $s_{-}^r(q) <0, \, \forall q$. 
(b) Bifurcation diagram in the $(\gamma_{\mathrm{a}}, \delta_{\mathrm{a}})$
plane. The yellow (\emph{resp.} blue) domain corresponds to a stable
(\emph{resp.} unstable) homogeneous state. 
}
\end{figure}

\subsection{Summary}
\label{sec:model:summary}

To summarize Eqs.~(\ref{eq:cons:matter}-\ref{eq:friction}) and 
(\ref{eq:consteqsig}-\ref{eq:consteqp}), 
the dynamics is governed by a set of three coupled partial
differential equations, with periodic boundary conditions on all fields:  
\begin{eqnarray}
  \label{eq:dynrho}
\partial_t \rho + \partial_x (\rho v)   &=& 0
\\
  \label{eq:dynsig}
- \partial_x \pi + \eta \; \partial^2_x v  
  + \gamma_{\mathrm{a}} \; \partial_x \rho 
 &=&  \xi \, v - t_{\mathrm{a}} \, p     \\
  \label{eq:dynp}
\partial_t p + v \, \partial_x p 
+ \lambda  \; p \, \partial_x v 
+ \alpha_{\mathrm{a}}  \; p \, \partial_x p
   &=& 
\Gamma_{\mathrm{p}} h +  \delta_{\mathrm{a}} \; \partial_x \rho 
\end{eqnarray}
where  $h$ and $\pi$ are defined by \eqref{eq:val:h} and 
\eqref{eq:val:pi} respectively.

\begin{figure}[!t]
\centering
\showfigures{
(a)\includegraphics[scale=0.25]{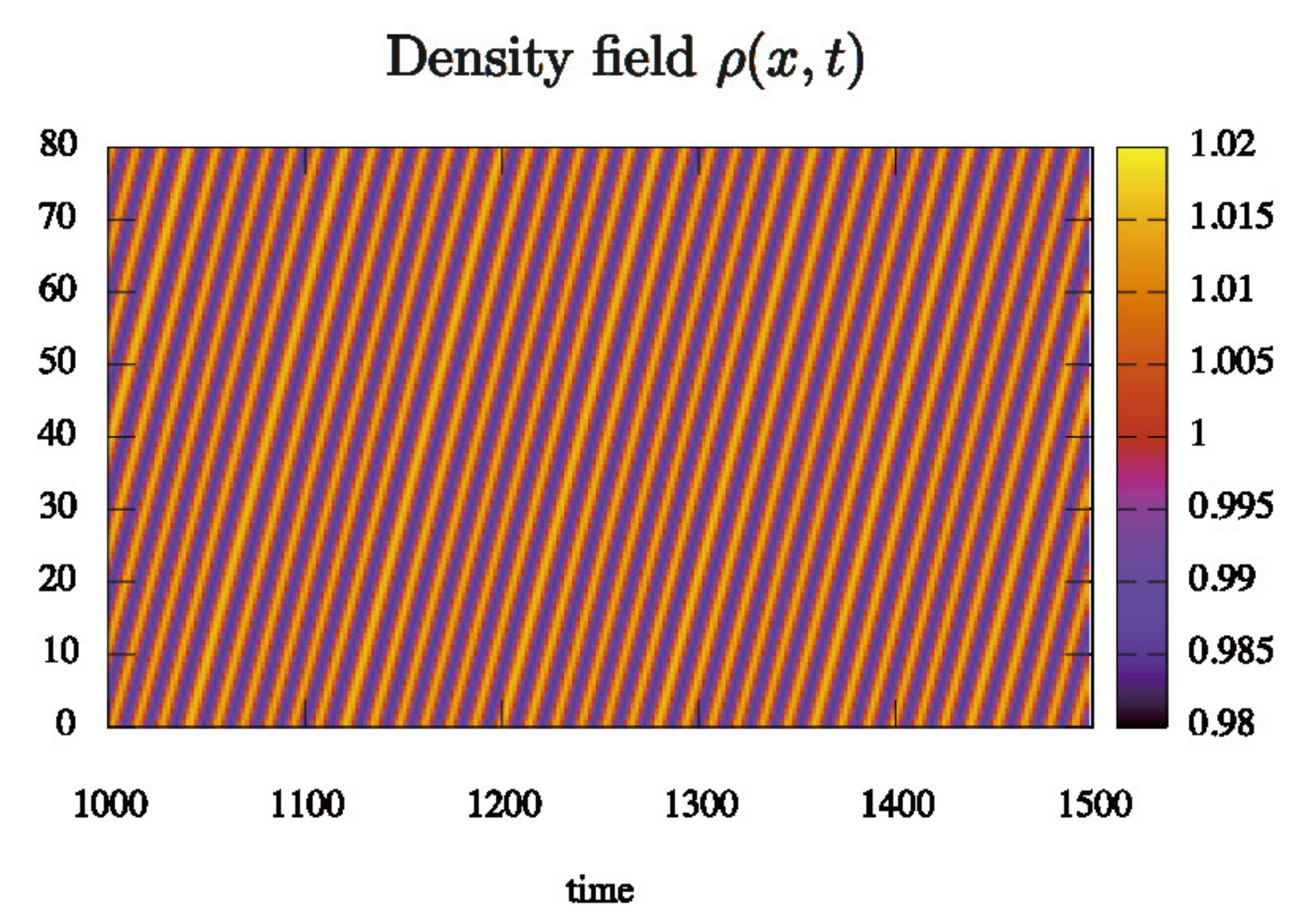}
\vspace*{0.1cm}
(b)\includegraphics[scale=0.25]{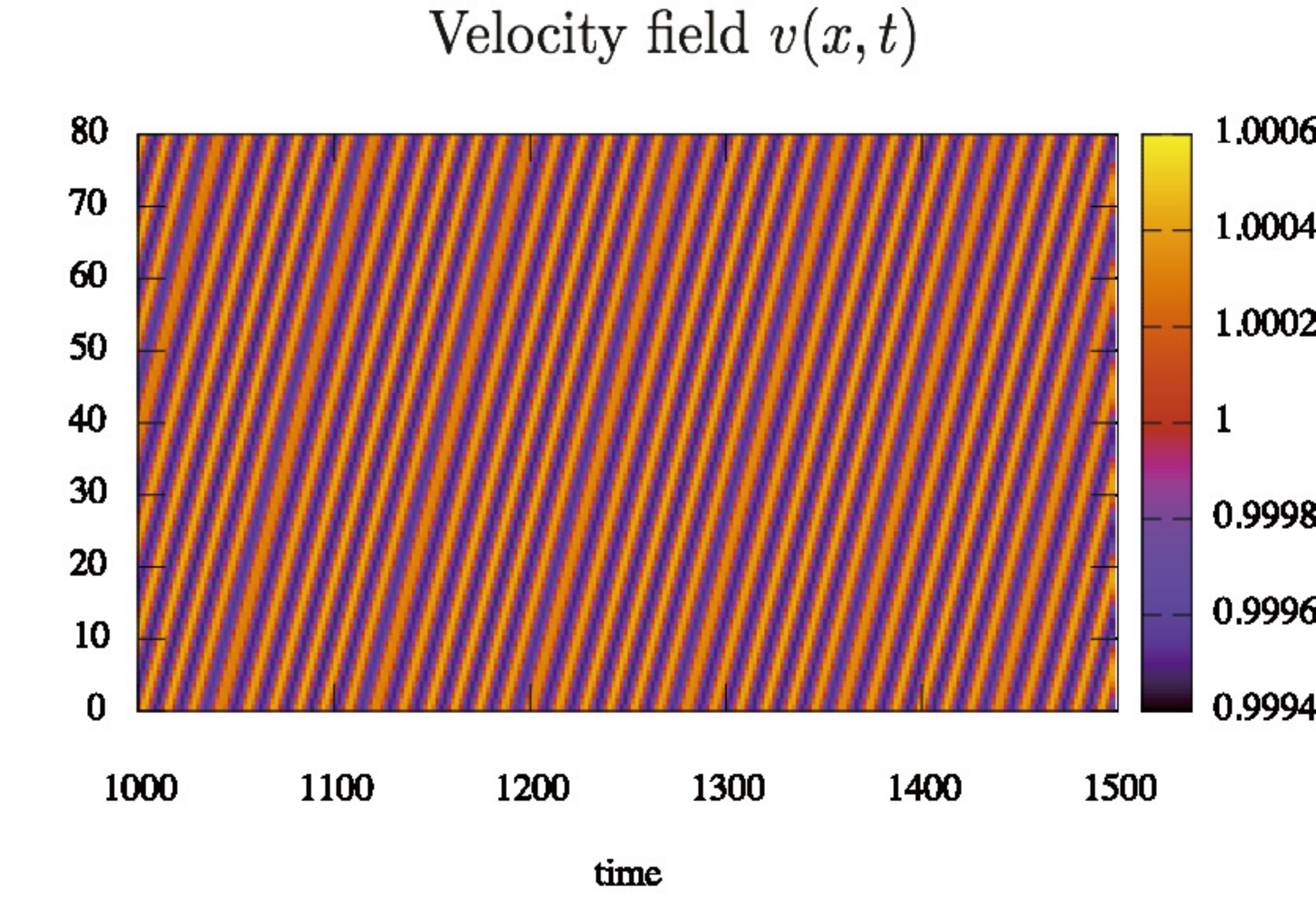}
}
\caption{\label{fig:num:wave}
\textbf{A traveling wave:} 
all parameters are set equal to one $K=\rhoe=a_{2}=\eta=\lambda=\Gamma_{p}=
t_{\mathrm{a}}=\alpha_{\mathrm{a}}=\delta_{\mathrm{a}}=\gamma_{\mathrm{a}}=1$,
except for $\xi = p_{0}=0.5$ and $K_{4}=0.0005$.
The cell density (a) and velocity (b) kymographs are presented when 
$t \ge 1000$, after transients have died out. The polarity and stress fields 
also exhibit stable traveling wave solutions for the same parameter values 
(not shown).
}
\end{figure}

\section{Linear stability analysis}
\label{sec:model:stab}

The system (\ref{eq:dynrho}-\ref{eq:dynp}), supplemented with 
(\ref{eq:val:h}-\ref{eq:val:pi}) and (\ref{eq:def:fp}-\ref{eq:def:frho})
admits  stationary, homogeneous solutions
$$(\rho, p, v) = (\rho_0, p_0, v_0)$$ 
with finite polarity  $p_0 = \sqrt{a_2/a_4}$, finite velocity
\mbox{$v_0 = t_{\mathrm{a}} p_0/\xi$}, and arbitrary 
$\rho_0$. For simplicity, we set $\rho_0 = \rhoe$ in the following.

We study the growth rate $s$ of perturbations of wave number $q$:
\begin{equation}
  \label{eq:linear:Ansatz} 
(\rho, p,  v) = (\rhoe, p_0,  v_0) + ( \delta \rho, \delta p, \delta v) \,
e^{s t - i q x} \,.
\end{equation}
Substituting \eqref{eq:linear:Ansatz} into (\ref{eq:dynrho}-\ref{eq:dynp})
with (\ref{eq:val:h}-\ref{eq:val:pi})
and (\ref{eq:def:fp}-\ref{eq:def:frho}),
we obtain to linear order a set of coupled equations, written in matrix form 
\begin{equation}
\label{eq:eq:A}
\mathcal{L} 
\left( \begin{array}{c}
\delta \rho  \\
\delta p  \\
\delta v
\end{array} \right)
=
\left( \begin{array}{c}
0  \\
0  \\
0
\end{array} \right)
\end{equation}
with 
\pagebreak
\begin{widetext}
\begin{equation*}
\label{eq:def:A}
\mathcal{L} =
\left( \begin{array}{ccc}
s -i q v_0 & 0  & -i q \rhoe \\ 
i q  \left( \frac{1}{K \rhoe} - \gamma_{\mathrm{a}} + \nu \rhoe q^4 \right) &
t_{\mathrm{a}}  & -(\xi + \eta q^2) \\
i q \delta_{\mathrm{a}} & s + 2 a_2 \Gamma_{\mathrm{p}}  
- i q \left( v_0 + \alpha_{\mathrm{a}} p_0 \right) &
- i q \lambda p_0 
\end{array} \right)
\end{equation*}
\end{widetext}
whose solution yields the wavenumber dependence
of the complex-valued growth rates $s_{\pm}(q) = s_{\pm}^r(q) + i s_{\pm}^i(q)$.
A Hopf bifurcation occurs above a threshold value of a control parameter
where the real part of the growth rate $s_{+}^r$ becomes positive, 
while its imaginary part is non-zero $s_{+}^i \neq 0$.
For parameter values above threshold, the velocity of traveling waves 
is given by $c =  s_{+}^i(q_{\mathrm{max}})/q_{\mathrm{max}}$, 
where $q_{\mathrm{max}}$ is the most unstable wavenumber 
$q_{\mathrm{max}} = \mathrm{argmax}_q(s_{+}^r(q))$.

We performed numerically the linear stability analysis.
In Fig.~\ref{fig:all:one}a, we plot $s_{+}^r(q)$ 
with all parameter values set equal to one ($t_{\mathrm{a}} = \eta = 
\Gamma_{\mathrm{p}} = \lambda = K = \rhoe = \alpha_{\mathrm{a}} = 
\delta_{\mathrm{a}} = \gamma_{\mathrm{a}}  = 1$), except $\xi = p_0 = 0.5$, 
and find that traveling waves are expected in this system, 
with a wave propagation velocity equal to $c = c(q_{\mathrm{max}}) = 1.07$, 
larger than the average velocity of the flow $v_0 = 1$.
In Fig.~\ref{fig:all:one}b, we plot the bifurcation diagram in the
$(\gamma_{\mathrm{a}}, \delta_{\mathrm{a}})$ plane:
a Hopf bifurcation occurs for large enough values of these
active parameters.

From Eqs.~(\ref{eq:fbal}-\ref{eq:friction}), we deduce
$$
- i q \delta \sigma = \xi \delta v - t_{\mathrm{a}} \delta p \,,
$$
where the first term on the right hand side is dissipative.
Eliminating $\delta \rho$ from \eqref{eq:eq:A}, we find that,
in the linear regime, 
$\delta p = (g^r(q) + i g^i(q)) \, \delta v$, where $g^r$ and $g^i$ are real
functions of the wavenumber, and obtain by substitution
$$
- i q \delta \sigma = 
\left( (\xi - t_{\mathrm{a}} g^r(q)) - i t_{\mathrm{a}} g^i(q) \right) \delta v  \,.
$$
The presence of an imaginary contribution $i t_{\mathrm{a}} g^i(q)$ 
may explain the apparently ``elastic'' behaviour of the mechanical waves 
as presented in \cite{Serra-Picamal2012}, while being in fact due to the
active and polar nature of this compressible, viscous material 
(see a similar discussion in \cite{Blanch2017b}).

\begin{figure}[!t]
\centering
\showfigures{
\includegraphics[scale=0.3]{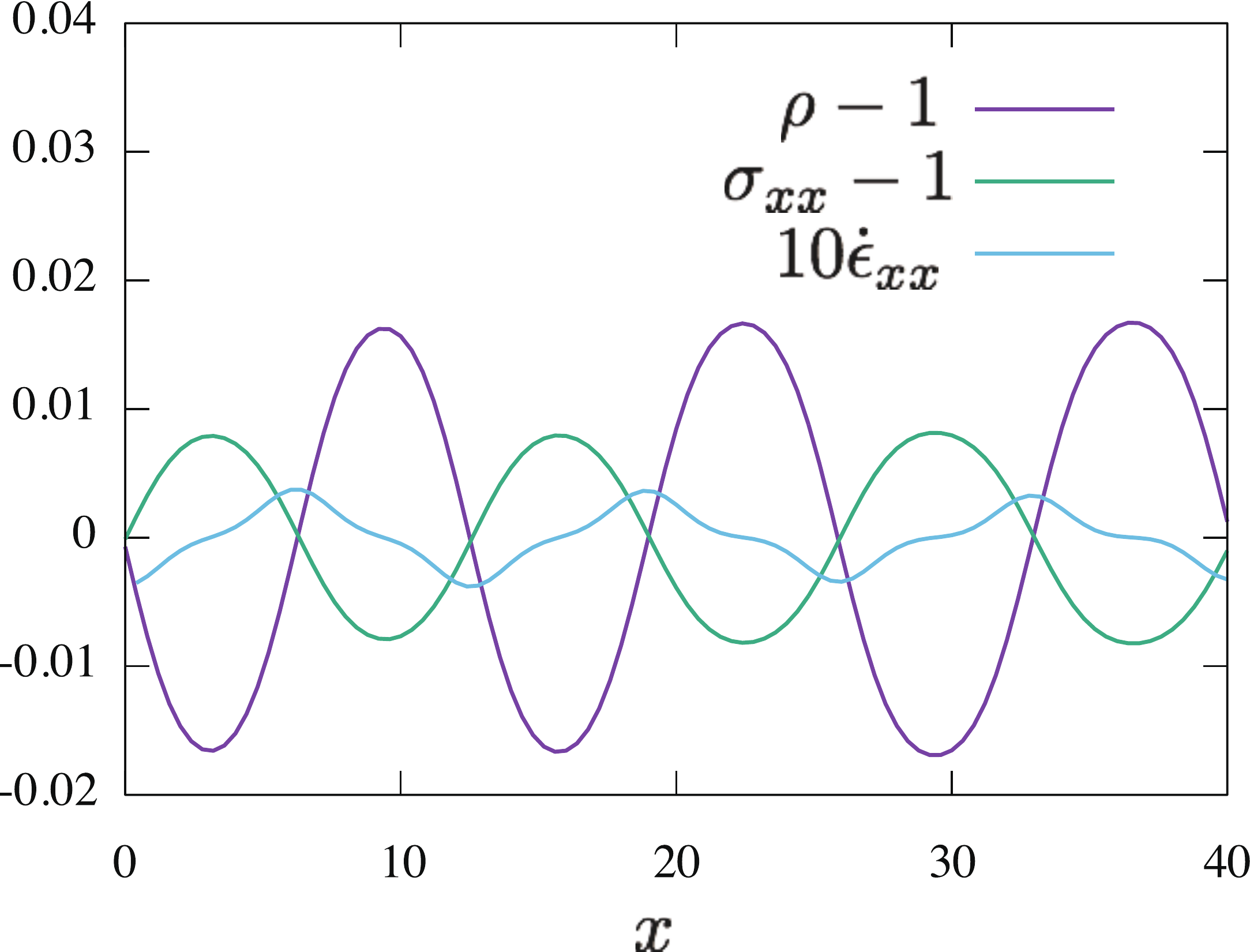}
}
\caption{\label{fig:1d-profiles}
\textbf{Spatial profiles} of the density, stress and  strain rate 
fields at $t=1000$. Parameter values are as in Fig.~\ref{fig:num:wave}.
}
\end{figure}

\section{Numerical simulations}
\label{sec:numerics}

In this section, we present the results of numerical simulations of 
Eqs.~(\ref{eq:dynrho}-\ref{eq:dynp}),
supplemented with (\ref{eq:val:h}-\ref{eq:def:frho}).
As an initial condition, 
we take the uniform state with $\rho=\rhoe$, $p=p_{0}$, 
$v=v_0 = t_{\mathrm{a}} p_{0}/\xi$ and add a noise of small amplitude.
We used a finite-difference method and integrated Eqs.~(\ref{eq:dynrho}-\ref{eq:dynp}) on a discretized 1d space with mesh size $\Delta x=0.01$ and time step $\Delta t=0.0001$ as follows: 
Given the density field $\rho(x,t)$ and the polarity field $p(x,t)$ at a time step $t$,  we solve Eq.~(\ref{eq:dynsig}) for the velocity field $v(x,t)$. 
Next, using Eq.~(\ref{eq:dynrho}) and Eq.~(\ref{eq:dynp}), we determine the density field $\rho(x,t+\Delta t)$ and the polarity field $p(x,t+\Delta t)$ at the next time step $t+\Delta t$ with the explicit Euler method.

In Fig.~\ref{fig:num:wave}, we set all parameters to one 
($K=\rhoe=a_{2}=\eta=\lambda=\Gamma_{p}=
\gamma_{\mathrm{a}}=t_{\mathrm{a}}=\alpha_{\mathrm{a}}=\delta_{\mathrm{a}}=1$)
except for $\xi = p_{0}=0.5$ and $K_{4}=0.0005$.  
As predicted by the linear stability analysis in the previous section, 
we confirm numerically  that the instability occurs and the 
propagation velocity $(c \sim 1.07)$ is slightly larger 
than the mean tissue velocity $v_0 = t_{\mathrm{a}} p_{0} /\xi=1$.
The transient time scale is of the order of $10^2$. 
In Fig.~\ref{fig:1d-profiles}, we show for the same parameter values 
the spatial profiles of the cell density, stress and  strain rate 
$\dot\epsilon_{xx}=\partial_{x} v(x,t)$ at $t=1000$. 
The stress and strain rate are out-of-phase, while the stress and the 
cell density are in antiphase.
If we consider that the cell area measured experimentally corresponds 
to the inverse of the cell density, both phase relations agree 
with experimental observations, which have been interpreted 
as evidence for an elastic rheology \cite{Serra-Picamal2012}. 
However, our constitutive equation \eqref{eq:consteqsig} is that of a 
compressible, viscous material, also endowed with active and 
polar properties.

\begin{figure}[!h]
\centering
\showfigures{
(a)\includegraphics[scale=0.25]{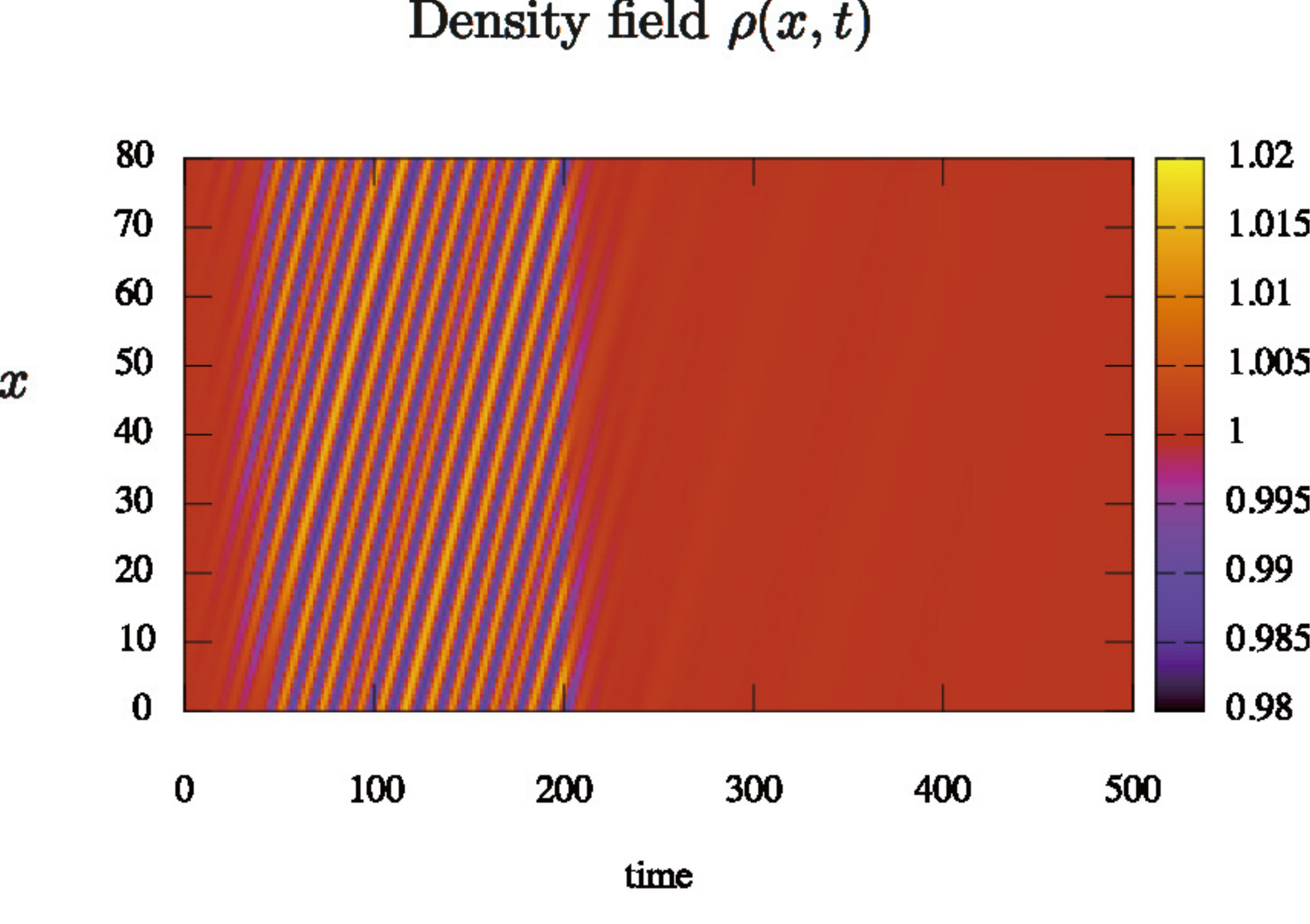}
\vspace*{0.1cm}
(b)\includegraphics[scale=0.25]{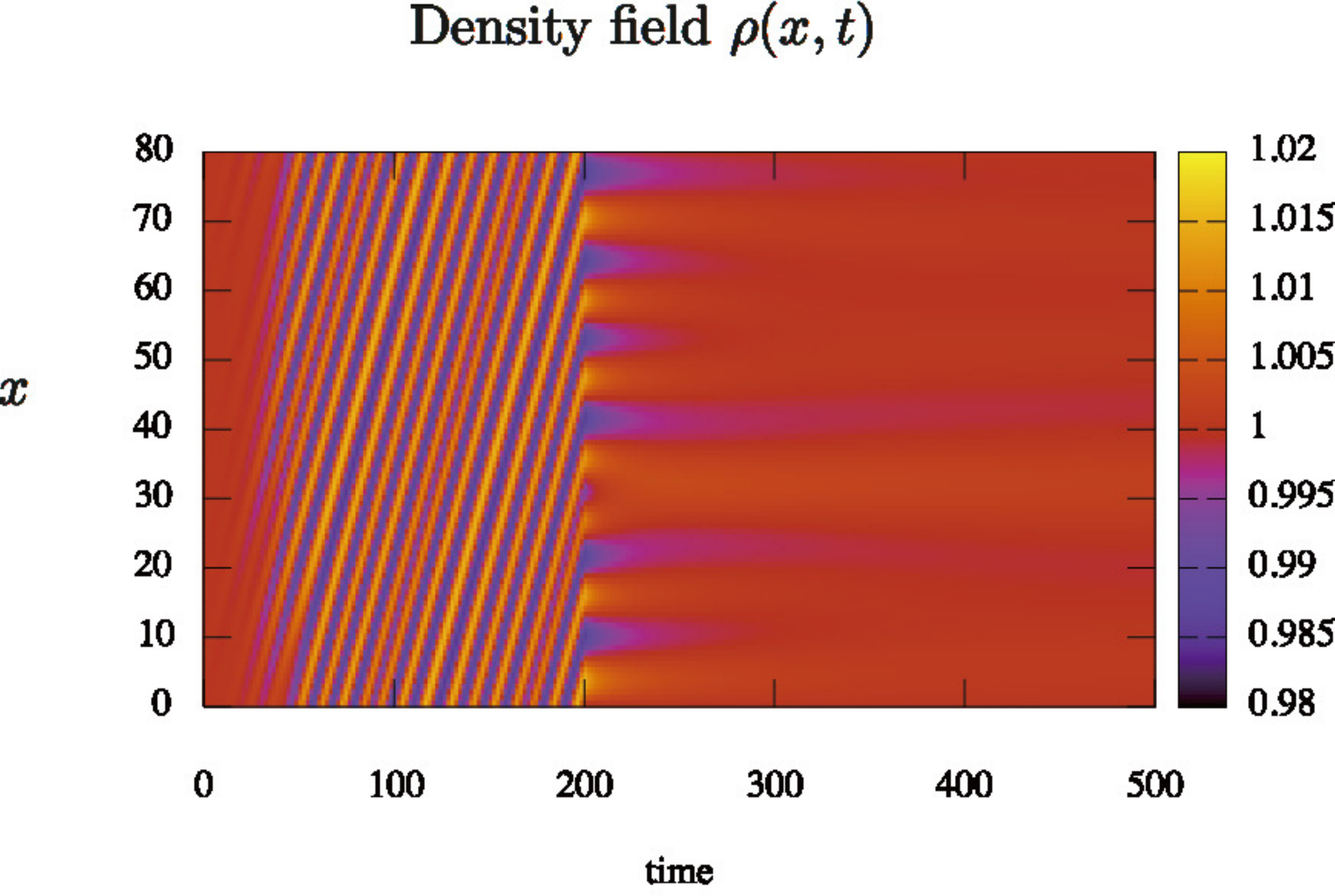}
}
\caption{\label{fig:num:inhibition}
\textbf{Numerical experiments mimicking inhibition assays:} 
density kymographs, with parameters as
in Fig.~\ref{fig:num:wave} until $t=200$.
(a) Experimental inhibition of contractility by blebbistatin is mimicked by
lowering the active parameters $\delta_{\mathrm{a}}$ and $\gamma_{\mathrm{a}}$
to $0.5$ when $t \ge 200$. 
(b) Experimental inhibition of Arp 2/3 is mimicked by
lowering $p_0$ to a large negative value $-500$ when $t \ge 200$. 
In both cases the traveling wave pattern disappears when $t \ge 200$, 
as predicted by linear stability analysis.
}
\end{figure}

Numerical simulations allow to mimic inhibition assays
by tuning the value(s) of parameter(s) that would be changed
due to the application of a drug. 
When we lower the value of the active parameters to
$\delta_{\mathrm{a}}=0.5$, $\gamma_{\mathrm{a}}=0.5$ 
at $t=200$, we observe that the traveling wave 
rapidly disappears (see Fig.~\ref{fig:num:inhibition}a), 
as observed experimentally when inhibiting contractility. 
In a similar manner, when we change the reference polarity value 
from a finite $p_0$ to $p_0 = 0$ (by setting  $a_{2}=-500$ at $t=200$),
the traveling waves also disappears quickly
(see Fig.~\ref{fig:num:inhibition}b), as observed experimentally 
when inhibiting Arp 2/3.

\begin{figure}[!h]
\centering
\showfigures{
(a)\includegraphics[scale=0.25]{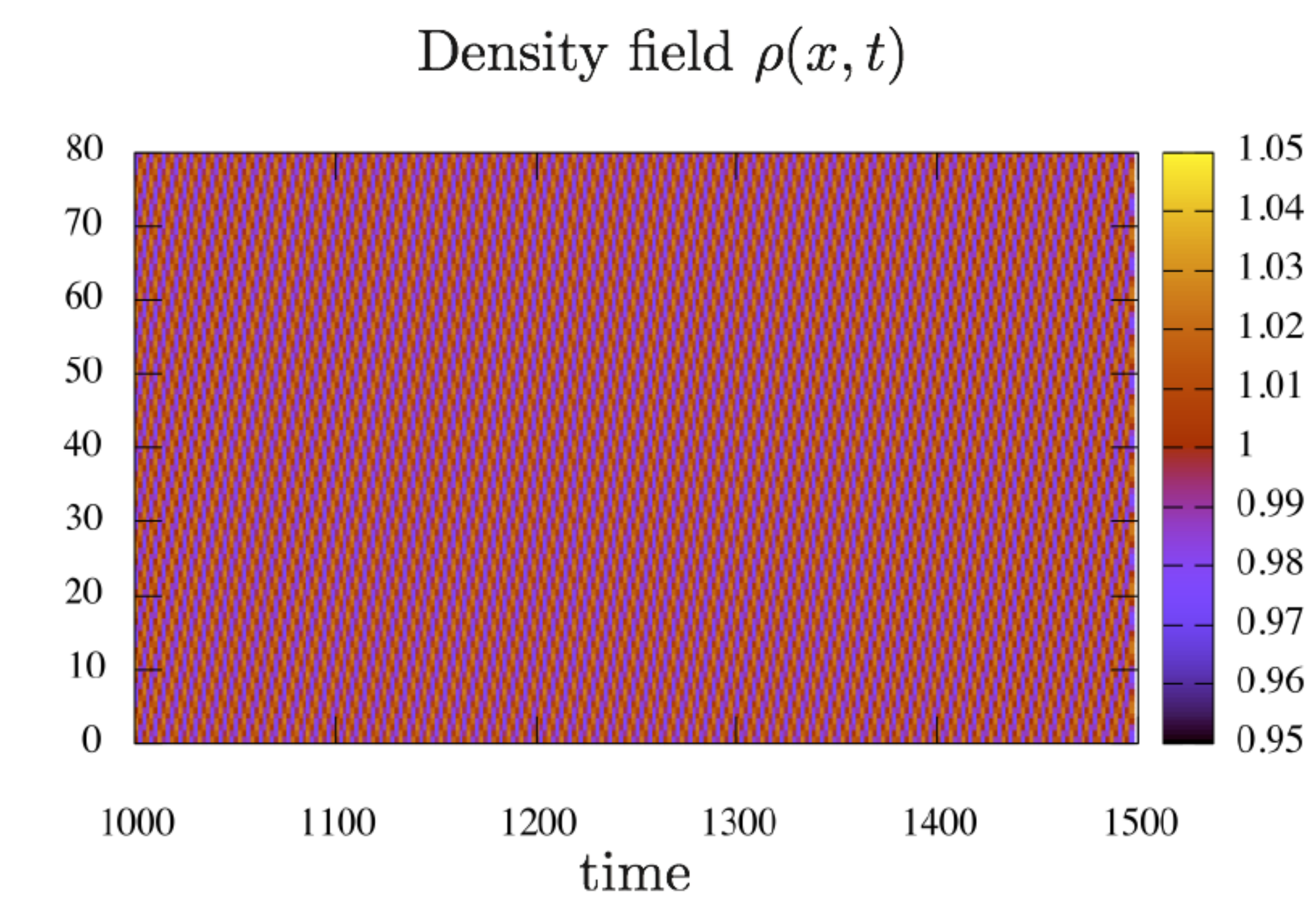}
\vspace*{0.1cm}
(b)\includegraphics[scale=0.25]{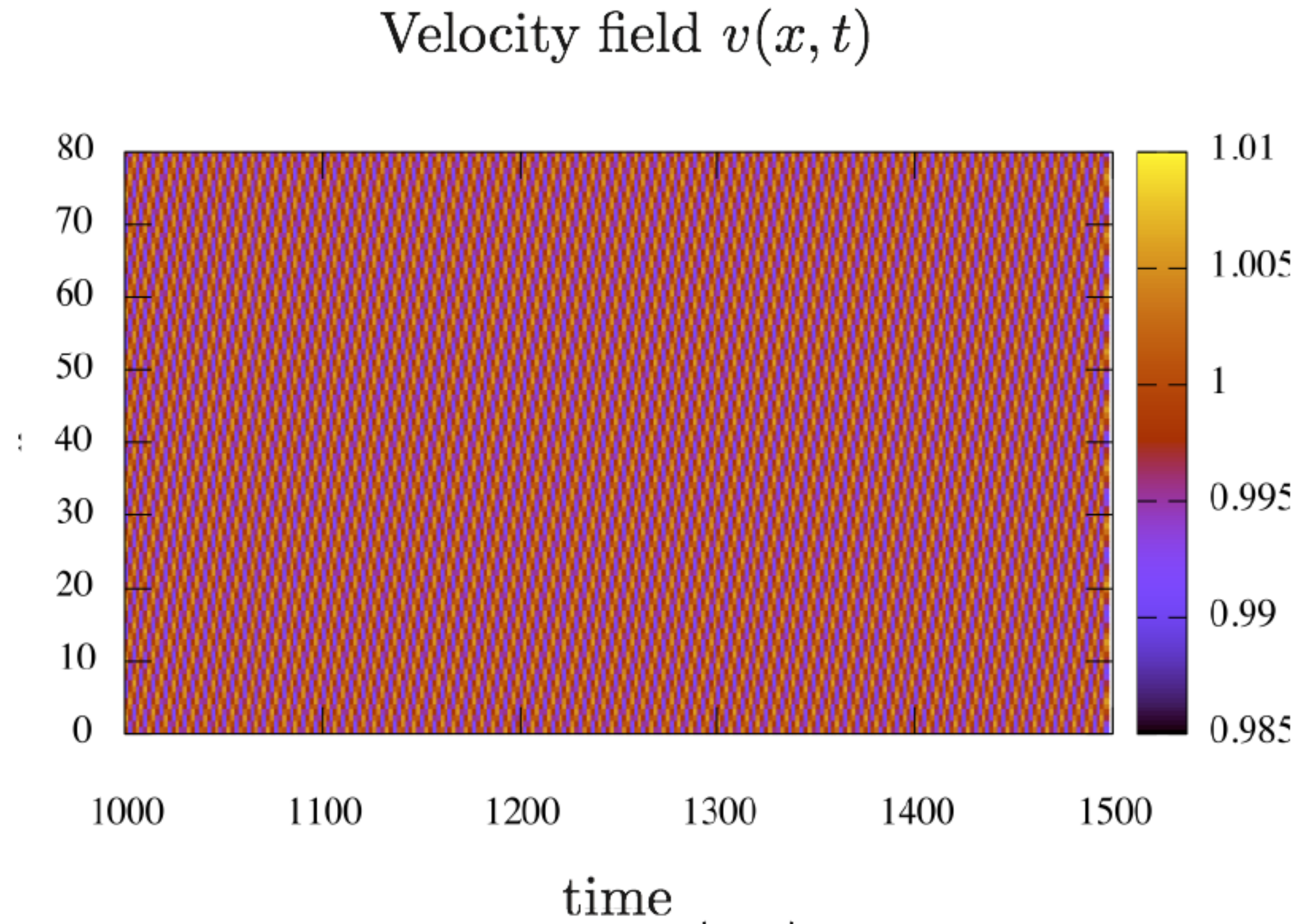}
}
\caption{\label{fig:num:wave:far}
\textbf{A traveling wave:} 
Parameters are as in Fig.~\ref{fig:num:wave}, except the active para\-meters
$\alpha_{\mathrm{a}} = 2$, $\gamma_{\mathrm{a}} = \delta_{\mathrm{a}} = 3$.
Kymographs of (a) the cell density (a); and (b) the velocity field  
are presented when $t \ge 1000$, after transients have died out. 
The polarity and stress fields also exhibit stable traveling 
wave solutions for the same parameter values (not shown).
}
\end{figure}

Finally, Fig.~\ref{fig:num:wave:far} shows an example of a traveling wave 
observed farther from threshold,  with $\alpha_{\mathrm{a}} = 2$,
$\gamma_{\mathrm{a}} = \delta_{\mathrm{a}} = 3$, and other parameters as in 
Fig.~\ref{fig:num:wave}. The propagation velocity  is accordingly larger 
$c \sim 1.3$.

\begin{figure}[!h]
\centering
\showfigures{
(a)\includegraphics[scale=0.45]{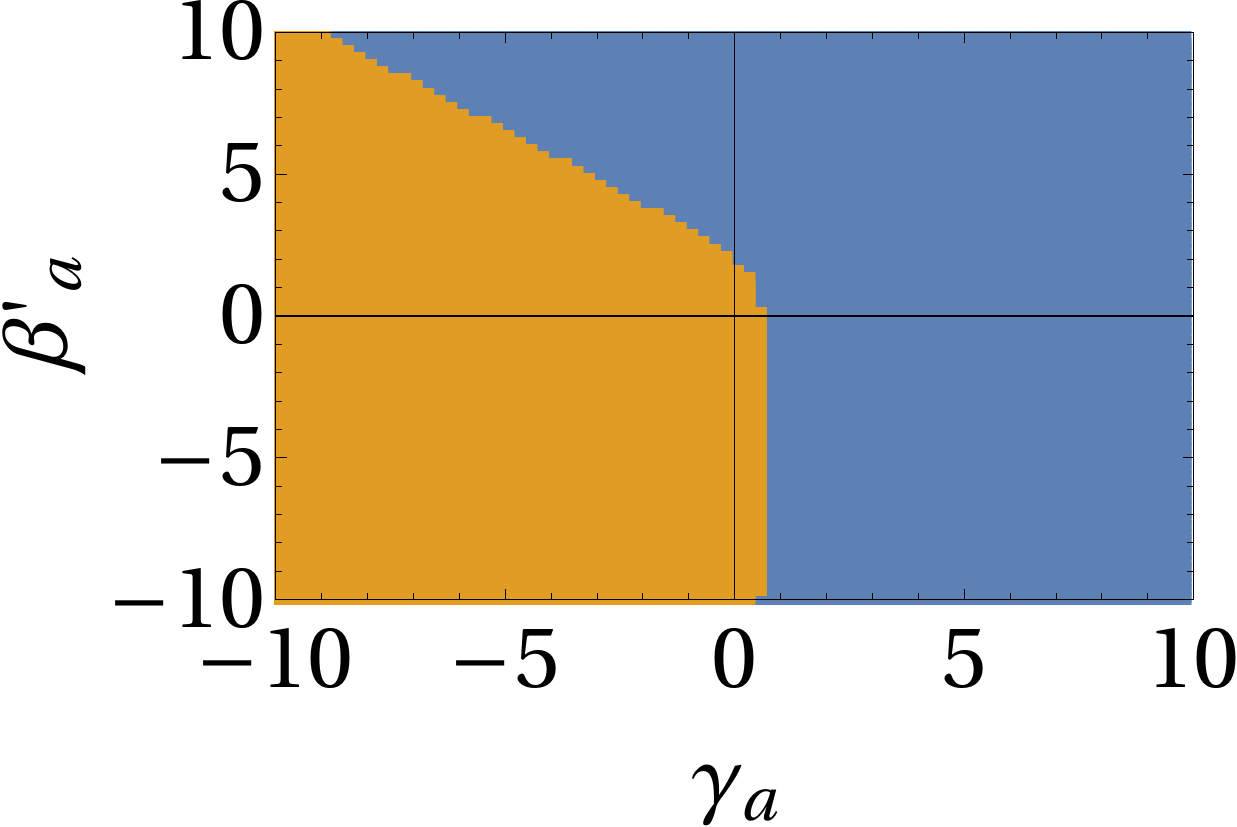}
\vspace*{0.1cm}
(b)\includegraphics[scale=0.45]{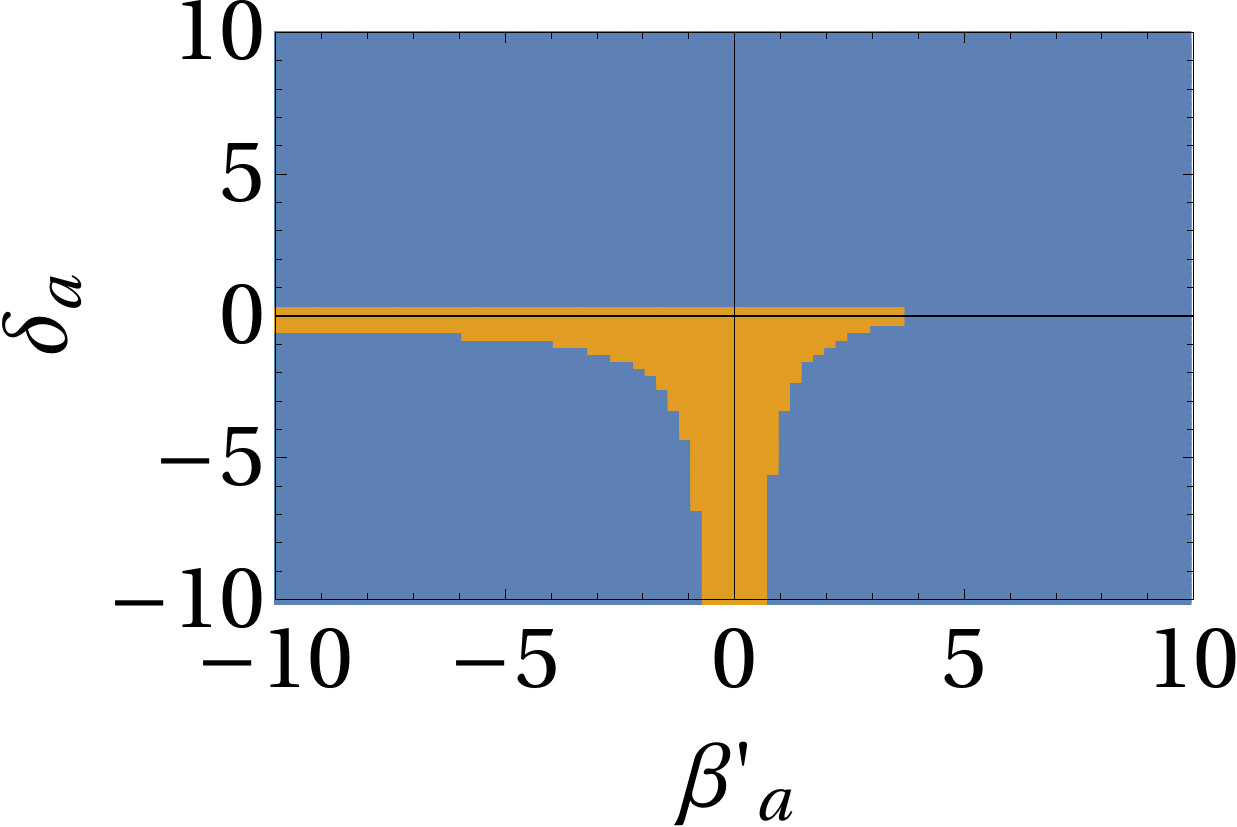}
}
\caption{\label{fig:betaprimea}
\textbf{Bifurcation diagrams:} 
(a) in the $(\gamma_{\mathrm{a}}, \beta'_{\mathrm{a}})$ plane, with 
$\delta_{\mathrm{a}} = 1$. (b) in the $(\beta'_{\mathrm{a}}, \delta_{\mathrm{a}})$  plane, with $\gamma_{\mathrm{a}} = 1$.  
Our minimal model is supplemented with
the $\beta'_{\mathrm{a}} \; p^2 $ term in \eqref{eq:add:sigma}, while}
other parameter values are as in Fig.~\ref{fig:all:one}b. 
The yellow (\emph{resp.} blue) domain corresponds to a stable
(\emph{resp.} unstable) homogeneous state. 
\end{figure}

\section{Discussion}
\label{sec:disc}

\subsection{Robustness}
\label{sec:disc:add}

Additional terms also invariant under simultaneous inversion of space 
and polarity are allowed by symmetry in the definition of the free energy
density \eqref{eq:def:f}. A more complete, but 
also more complex expression of the free energy density as an expansion in 
terms of cell density, polarity, and their gradients reads
\begin{align*}
  f = \, &   \frho(\rho) +
      \frac{\kr}{2} \; \left( \partial_x \rho \right)^2
      + \frac{\nu}{2} \, \left( \partial^2_x \rho \right)^2 \\
& + \fp(p) + \frac{\kf}{2} \; \left( \partial_x p \right)^2 
      + \frac{\nu_K}{2} \, \left( \partial^2_x p \right)^2\\
 & +      w_1  \, \partial_x p +  w_2 \, \rho  \, \partial_x p +
 w_3 \, p  \, \partial_x \rho +
 w_4 \, \partial_x \rho \, \partial^2_x p +
 w_5 \, \partial_x p \, \partial^2_x \rho + \ldots ,
\end{align*}
and is by no means exhaustive. It leads to more complex expressions 
of the conjugate fields $\pi$ and $h$. Here $\kf \ge 0$ is Frank's constant, 
the parameter $\nu_K \ge 0$ controls a term that further stabilizes large 
wavenumber modes, and the $w_i$'s couple density and polarity gradients. 
Note that the terms penalizing polarity gradients may
become relevant in the presence of topological defects. 

Similarly, additional nonlinear active terms could 
be included in the right hand side of \eqref{eq:consteqsig}:
\begin{equation}
\label{eq:add:sigma}
  \sigma  = -  \pi + \eta \; \partial_x v 
             + \beta_{\mathrm{a}} \; \partial_x p 
             + \beta'_{\mathrm{a}} \; p^2 
           + \sigma_{\mathrm{a}}  + \gamma_{\mathrm{a}} \, \rho + \ldots
\end{equation}
and of \eqref{eq:consteqp}:
\begin{equation}
\label{eq:add:pdot}
  \dot{p} = \Gamma_{\mathrm{p}} \; h + a_{\emph{a}}  \, p
 + a'_{\emph{a}} \, \rho  p
- \lambda  \; p \, \partial_x v 
 - \alpha_{\mathrm{a}}  \; p \, \partial_x p 
+ \delta_{\mathrm{a}}  \; \partial_x \rho
+ \ldots
\end{equation}
since they also respect the invariance properties of $\sigma$
and $\dot p$. 

We checked that these additional terms, while making the model 
more complex, do not change our qualitative conclusions: activity-driven 
traveling waves occur in large regions of the parameter space that 
govern active polar media (see also \cite{Giomi2012,Ramaswamy2016}).
As an example motivated by \cite{Blanch2017b}, we add to our minimal
model the $\beta'_{\mathrm{a}} \; p^2 $ term in \eqref{eq:add:sigma}, 
and show in Fig.~\ref{fig:betaprimea} the resulting bifurcation diagrams 
in the $(\gamma_{\mathrm{a}}, \beta'_{\mathrm{a}})$ plane, at constant 
$\delta_{\mathrm{a}}$ and in the $(\beta'_{\mathrm{a}}, \delta_{\mathrm{a}})$ 
plane, at constant $\gamma_{\mathrm{a}}$ (compare with 
Fig.~\ref{fig:all:one}b).

\begin{figure}[!t]
\showfigures{
(a)\includegraphics[scale=0.25]{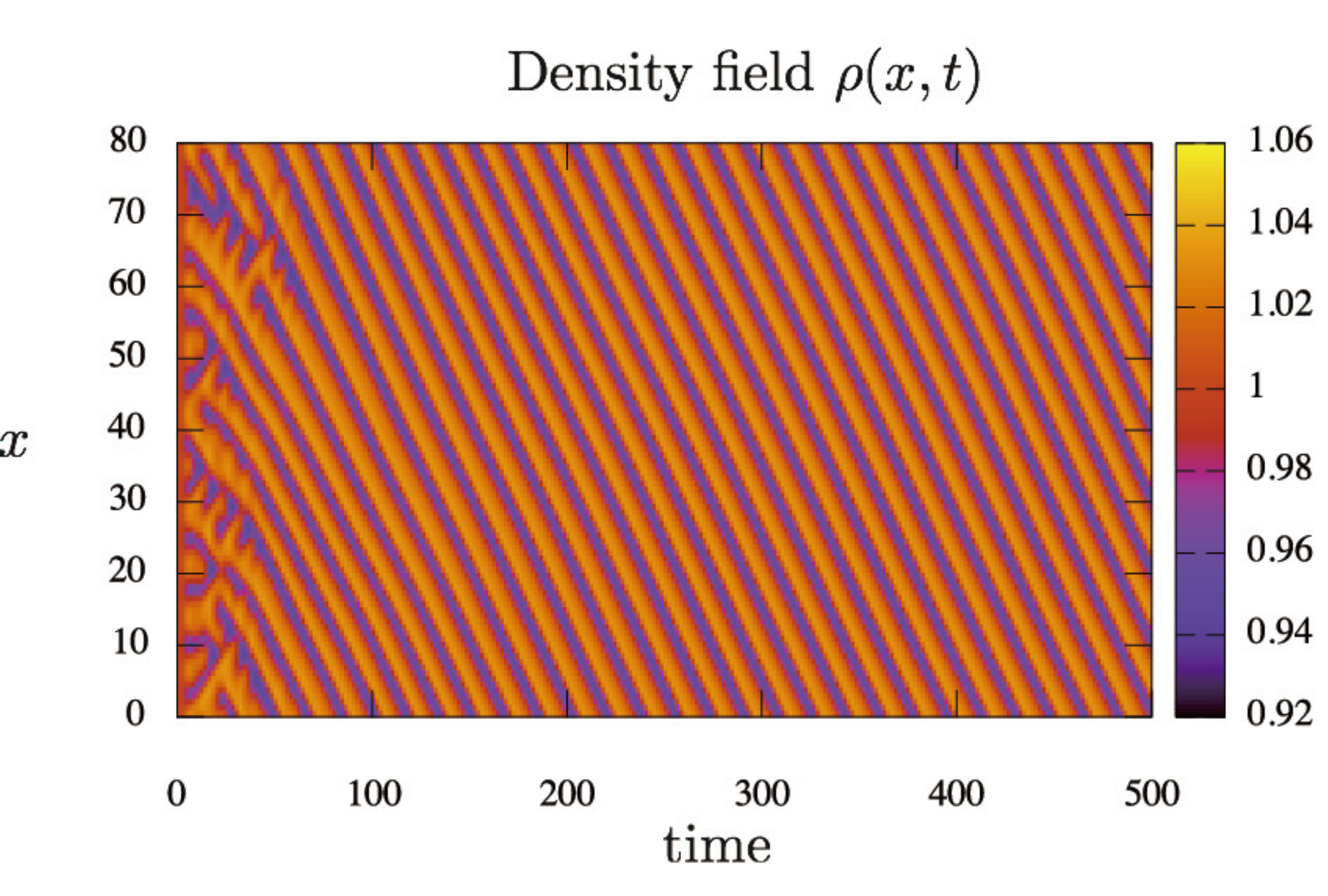}
\vspace*{0.1cm}
(b)\includegraphics[scale=0.25]{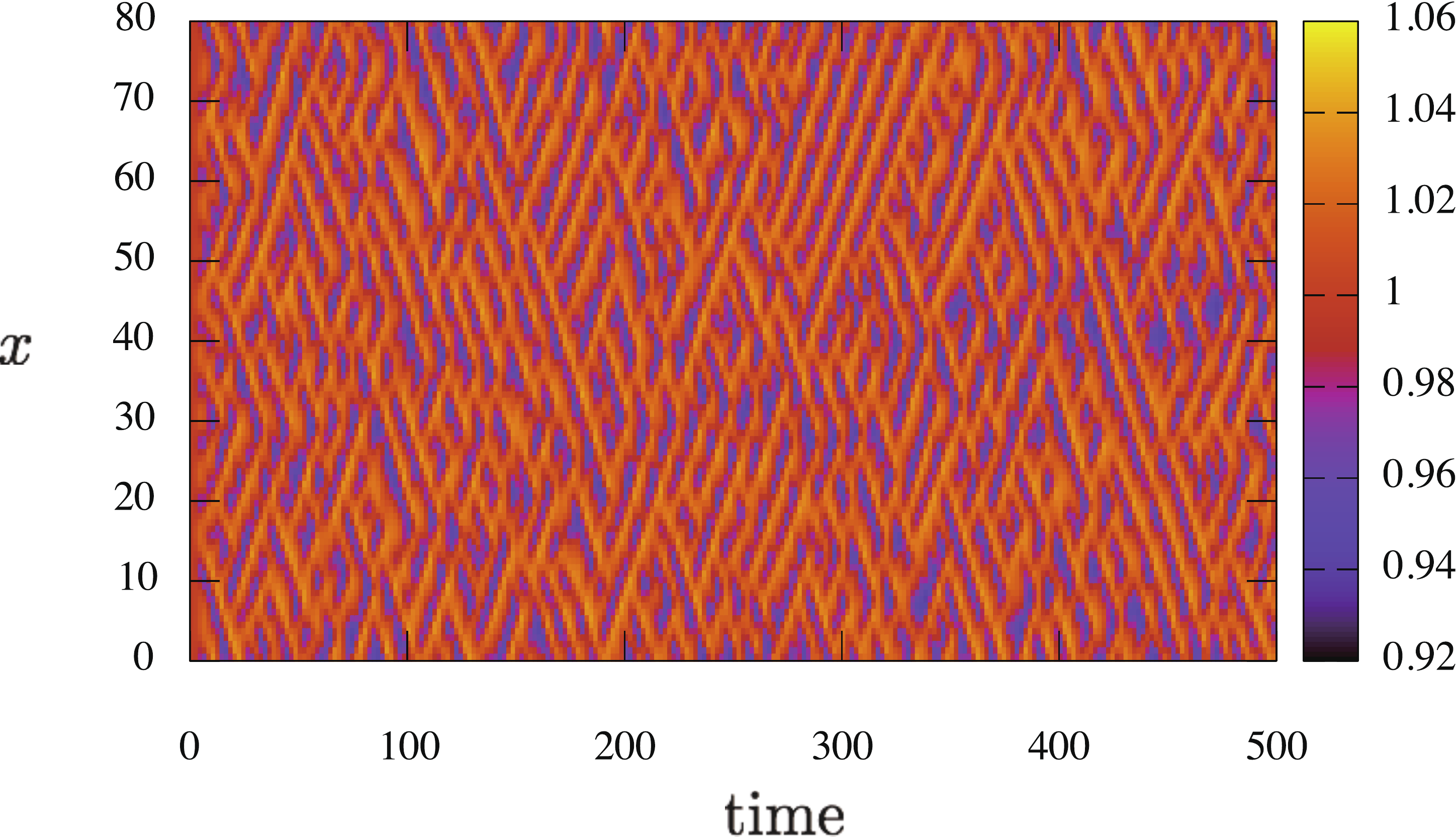}
\vspace*{0.1cm}
(c)\includegraphics[scale=0.25]{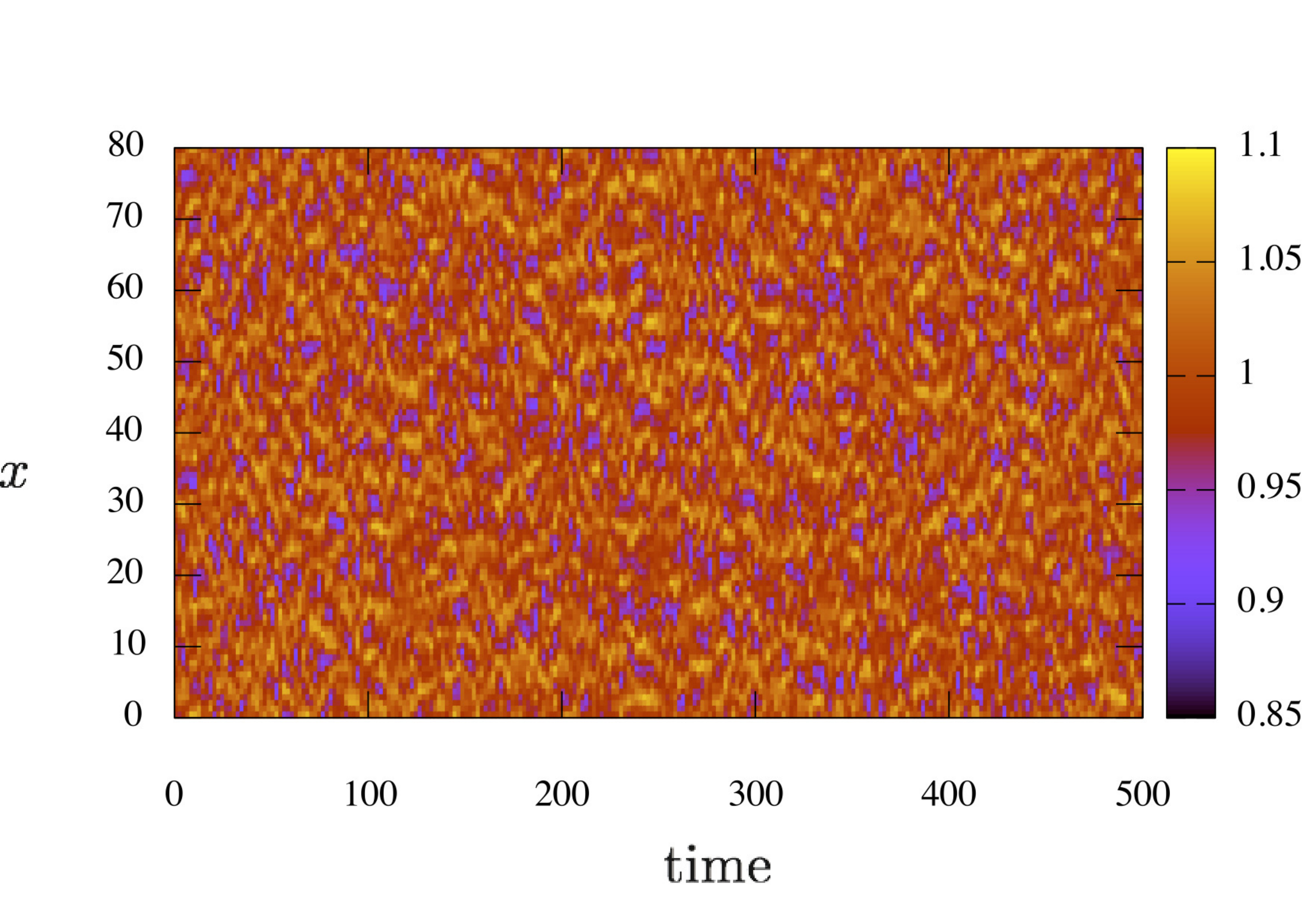}
}
\caption{\label{fig:num:wave:secondarybif}
\textbf{Secondary bifurcations.}
Our minimal model is supplemented with
the $\beta'_{\mathrm{a}} \; p^2 $ term in \eqref{eq:add:sigma}, with
parameters as in Fig.~\ref{fig:num:wave}, except $a_2=-1$, $\beta_{\mathrm{a}}=0.1$, $\gamma_{\mathrm{a}}$, $\delta_{\mathrm{a}}$ and $\beta'_{\mathrm{a}}$. 
Since $a_2<0$, the primary bifurcation is stationary. 
Density kymographs are shown, with  (a) a traveling wave:
$\gamma_{\mathrm{a}}=2$, $\delta_{\mathrm{a}} = 4$  and $\beta'_{\mathrm{a}}=20$;
(b) coexistence of rightwards and leftwards traveling waves:
$\gamma_{\mathrm{a}}=2$, $\delta_{\mathrm{a}} = 4$ and $\beta'_{\mathrm{a}}=30$; (c) a disordered pattern:
$\gamma_{\mathrm{a}}=4$, $\delta_{\mathrm{a}} = 8$ and $\beta'_{\mathrm{a}}=20$.
}
\end{figure}

\subsection{Behaviour close to $p = 0$}
\label{sec:disc:p00}

When the free energy is expanded close to $p = 0$, 
bifurcations driven by the active parameters 
$\gamma_{\mathrm{a}}$ and $\delta_{\mathrm{a}}$ also occur.
This case is obtained by setting $a_2<0$ in \eqref{eq:def:fp}, and
leads to substituting $p_0 \to 0$, $v_0 \to 0$ in (\ref{eq:eq:A}).
However, we find this time that $s_{+}^i(q_{\mathrm{max}}) = 0$,
\emph{i.e.} that the bifurcation is stationary and lead 
to \emph{standing waves}. This is consistent with previous work
on active polar materials \cite{Marcq2014}, and leads us to
expect that traveling waves may then arise farther from threshold 
due to secondary bifurcations. 

As an example, we show in Fig.~\ref{fig:num:wave:secondarybif}
that a secondary bifurcation occurs in the presence of 
$\beta_{\mathrm{a}}$ and $\beta'_{\mathrm{a}}$ : 
With the parameters $a_2=-1<0$, $\gamma_{\mathrm{a}} =2$, 
$\delta_{\mathrm{a}} = 4$, $\beta_{\mathrm{a}}=0.1$ and $\beta'_{\mathrm{a}}=20$, 
we found a travelling wave solution. 
When increasing $\beta'_{\mathrm{a}} $ and fixing the other parameters, 
we observed that the stationary periodic pattern becomes unstable for 
$\beta'_{\mathrm{a}} \gtrapprox 11.5$. We also  obtained disordered patterns by 
further increasing the active parameters. The cases shown in 
Fig.~\ref{fig:num:wave:secondarybif} should be considered as examples.
A full investigation of the complex spatio-temporal patterns generated by 
our model is left for future study.

\subsection{Model comparison}
\label{sec:disc:lit}

A significant difference with the treatments proposed in 
\cite{Banerjee2015,Notbohm2016} is that we take into account the  
compressibility of the monolayer \cite{Zehnder2015,Zehnder2015a} 
through the cell density balance equation 
(see \cite{Blanch2017b} for an alternative way to treat incompressibility).
Our formulation is more parsimonious than \cite{Banerjee2015,Notbohm2016} 
since we do not need to introduce an evolution equation for an 
additional chemical field in order to generate the instability.  
While the phenomenological model of wave generation proposed in 
\cite{Tlili2016} is based on a mechanotransduction pathway involving 
the protein Merlin \cite{Das2015}, other authors favour the ERK Map kinase 
as a possible mechanochemical mediator \cite{Notbohm2016}.
Our approach does not depend upon a specific biochemical mechanism. 
It is based on general physical principles, notably symmetry arguments,
which remain valid irrespective of the specific molecular mechanism
at play. We propose a mechanism where activity couples to the cell 
density field. Only detailed comparison with experimental data is 
liable to determine which of the possible nonlinearities is involved. 

Traveling waves are driven by either strong enough active stresses
or strong enough active couplings in the evolution equation of the 
polarity field. As a consequence, their emergence does 
not depend on a specific rheology. Due to the 1D geometry, our choice of 
a rheology is somewhat ambiguous. According to Eq.~(\ref{eq:consteqsig}), 
we assume that the epithelial rheology is that of a compressible, viscous 
fluid. Since it cancels at a finite value of the density, the pressure 
can also be interpreted in 1D as minus the elastic stress, 
$\pi = -\sigma_{\mathrm{el}}$. From this viewpoint, the epithelium rather 
behaves as a viscoelastic solid. An elastic behaviour corresponds to 
the limit of zero viscosity. Setting $\eta = 0$ only shifts 
the instability threshold without altering the structure of the 
bifurcation diagram. Thus traveling waves arise in our model 
for a viscous, an elastic, and a viscoelastic rheology (Kelvin model). 
We conjecture that the equations one would write for a
viscoelatic liquid (Maxwell model) would also give rise to 
activity-driven traveling waves \cite{Marcq2014}.

\section{Conclusion}
\label{sec:conc}

Within the framework of linear nonequilibrium thermodynamics, we wrote 
the constitutive equations for a one-dimensional, compressible, polar 
and active material, including lowest-order nonlinear active terms.
We showed that Hopf bifurcations occur at threshold values of active 
parameters, leading to traveling waves for the relevant mechanical fields
(density, velocity, polarity and stress).
We thus formulated and justified a minimal, physical
model of a tissue able to sustain traveling mechanical waves.
A lower value of active parameters may suppress the waves, in agreement
with experimental observation following the application of a drug inhibiting
contractility. Switching to a state without a preferred polarity 
may also suppress the waves, just as inhibiting Arp 2/3 does in the
cell monolayer. Not all epithelial cell types self-organize to 
generate traveling waves \cite{Vincent2015}: in the light of our model,
this suggests that active parameters are strongly cell-type dependent,
and may accordingly be above or below the instability threshold.

In order to describe faithfully the experimental
observations \cite{Serra-Picamal2012,Tlili2016}, the model needs 
to be extended to a 2D geometry, and to include a moving free boundary 
in the numerical simulations, close to which polar order may be confined 
within a band of finite extension \cite{Blanch2017a}. The same 
epithelial monolayers exhibit topological defects of the cell shape 
tensor field characteristic of a \emph{nematic} material \cite{Saw2017}. 
A more complete theory should take into account both types of 
orientational order.

\section*{Acknowledgements}

We wish to thank Carles Blanch-Mercader, Estelle Gauquelin, 
Fran\c{c}ois Graner, Shreyansh Jain, Benoit Ladoux and Sham Tlili 
for enlightening discussions and helpful suggestions. S.Y. was supported 
by Grant-in-Aid for Young  Scientists a(B) (15K17737), 
Grants-in-Aid for Japan Society for the Promotion of Science 
(JSPS) Fellows (Grant No. 263111), and the JSPS Core-to-Core Program 
"Non-equilibrium dynamics of soft matter and information".


\providecommand*{\mcitethebibliography}{\thebibliography}
\csname @ifundefined\endcsname{endmcitethebibliography}
{\let\endmcitethebibliography\endthebibliography}{}

\end{document}